\documentclass[twocolumn]{aastex631}

\shorttitle{Planetary Engulfment Prognosis}
\shortauthors{Stephen R. Kane}

\begin{document}

\title{Planetary Engulfment Prognosis within the $\rho$ CrB System}

\author[0000-0002-7084-0529]{Stephen R. Kane}
\affiliation{Department of Earth and Planetary Sciences, University of
  California, Riverside, CA 92521, USA}
\email{skane@ucr.edu}

%%%%%%%%%%%%%%%%%%%%%%%%%%%%%%%%%%%%%%%%%%%%%%%%%%%%%%%%%%%%%%%%%%%%

\begin{abstract}

Exoplanets have been detected around stars at various stages of their
lives, ranging from young stars emerging from formation, to latter
stages of evolution, including white dwarfs and neutron stars. Post
main sequence stellar evolution can result in dramatic, and
occasionally traumatic, alterations to the planetary system
architecture, such as tidal disruption of planets and engulfment by
the host star. The $\rho$~CrB system is a particularly interesting
case of advanced main sequence evolution, due to the relative late age
and brightness of the host star, its similarity to solar properties,
and the harboring of four known planets. Here, we use stellar
evolution models to estimate the expected trajectory of the $\rho$~CrB
stellar properties, especially over the coming 1.0--1.5 billion years
as it evolves off the main sequence. We show that the inner three
planets (e, b, and c) are engulfed during the red giant phase and
asymptotic giant branch, likely destroying those planets via either
evaporation or tidal disruption at the fluid body Roche limit. The
outer planet, planet d, is briefly engulfed by the star several times
toward the end of the asymptotic giant branch, but the stellar mass
loss and subsequent changing planetary orbit may allow the survival of
the planet into the white dwarf phase of the stellar evolution. We
discuss the implications of this outcome for similar systems, and
describe the consequences for planets that may lie within the
Habitable Zone of the system.

\end{abstract}


%%%%%%%%%%%%%%%%%%%%%%%%%%%%%%%%%%%%%%%%%%%%%%%%%%%%%%%%%%%%%%%%%%%%

\section{Introduction}
\label{intro}

From the plethora of exoplanet discoveries, a vast array of system
architectures have been revealed, many of which differ significantly
from the solar system
\citep{ford2014,winn2015,horner2020b,kane2021d,mishra2023a,mishra2023b}.
The majority of these exoplanet discoveries have occurred through the
use of the transit or radial velocity (RV) methods. Although the
stellar target sample is dominated by main sequence (MS) stars, there
are numerous surveys that have focused their efforts on evolved stars,
such as sub-giant and giant stars
\citep[e.g.,][]{hekker2007,wittenmyer2015d,jeong2018b}. There are
several prominent examples of exoplanets around giant stars, such as
the highly eccentric planet orbiting iota Draconis
\citep{frink2002,kane2010a,hill2021,campante2023} and the planetary
companion to Pollux \citep{hatzes2006,reffert2006b}. Such giant star
system architectures are of particular interest with respect to the
effects of post-MS evolution
\citep{villaver2009,veras2016a,macleod2018b,rapoport2021}. Predicting
the fate of the planets in these systems is of interest from a stellar
evolution, orbital evolution, and planetary habitability
perspective. The predicted effect of giant star engulfment of a
substellar companion can vary depending on the mass of the host star
and companion \citep{hon2023}. For example, engulfed giant planets may
experience substantial drag, resulting in orbital decay and eventual
tidal disruption at the Roche limit well interior to the stellar
radius \citep{oconnor2023a}. On the other hand, brown dwarf companions
may survive the red giant phase and continue to orbit the subsequent
white dwarf host relatively unscathed \citep{maxted2006}. Moreover,
the Habitable Zone (HZ) of the host star
\citep{kasting1993a,kane2012a,kopparapu2013a,kopparapu2014,kane2016c,hill2018,hill2023}
is profoundly affected by the change in luminosity and effective
temperature that occurs with post-MS evolution \citep{ramirez2016a}
with corresponding dramatic effects for HZ terrestrial planets
\citep{lopez2005,vonbloh2009,kozakis2019}. Thus, old MS stars are
useful templates from which to consider the consequences of stellar
evolution on the harbored planets.

One of the earliest exoplanet discoveries was that of the Rho Coronae
Borealis (HD~143761, HIP~78459, hereafter $\rho$~CrB)
system. $\rho$~CrB is a bright ($V = 5.39$) and nearby ($d =
17.51$~pc) MS star, with a spectral type of G0. Although the star has
properties similar to solar, it is $\sim$5\% less massive, $\sim$30\%
larger, and has sub-solar metallicity
\citep{santos2003a,takeda2007a,vonbraun2014,rosenthal2021,brewer2023}. Age
estimates for the star have yielded consistently high values, making
$\rho$~CrB one of the nearest solar-type stars with close proximity to
evolving off the MS \citep{metcalfe2021}. The planetary system
associated with the star was first discovered by \citet{noyes1997} via
the RV detection of a giant planet orbiting with a period of
$\sim$40~days. Subsequent RV monitoring by \citet{fulton2016} revealed
a second planet in the system with an orbital period of
$\sim$100~days. More recent RV efforts have focused on extreme
precision RV (EPRV) methods, pushing RV detection efficiency to
significantly smaller planetary masses and longer orbital periods than
previous instruments were capable of achieving
\citep{fischer2016}. One such instrument, the EXtreme-PREcision
Spectrograph (EXPRES) \citep{blackman2020c,petersburg2020}, monitored
$\rho$~CrB and discovered two more planets with orbital periods of
$\sim$13~days and $\sim$281~days, raising the total planet inventory
for the system to four \citep{brewer2023}. The relative proximity of
the $\rho$~CrB system, the even spacing of the known planetary orbits,
the sub-solar metallicity of the host star and its large age, make the
system an interesting case study regarding the aftermath of MS
evolution for the planets in the system.

Here, we present a stellar evolution model for $\rho$~CrB in the
context of the planetary orbits and the effect of post MS evolution on
their survivability. Section~\ref{arch} describes the stellar and
planetary parameters, and the system architecture. Section~\ref{mist}
provides details of the stellar evolution model, and the modifications
to the fundamental stellar parameters as the star evolves off the
MS. Section~\ref{prog} overlays the stellar evolution model on the
system architecture, quantifying the limits whereby specific planets
are engulfed by the host star. We discuss the implications of our
results in Section~\ref{discussion}, including outcomes for known
planets and potential HZ planets, and provide concluding remarks in
Section~\ref{conclusions}.

%%%%%%%%%%%%%%%%%%%%%%%%%%%%%%%%%%%%%%%%%%%%%%%%%%%%%%%%%%%%%%%%%%%%

\section{System Architecture}
\label{arch}

\begin{figure}
  \includegraphics[angle=270,width=8.5cm]{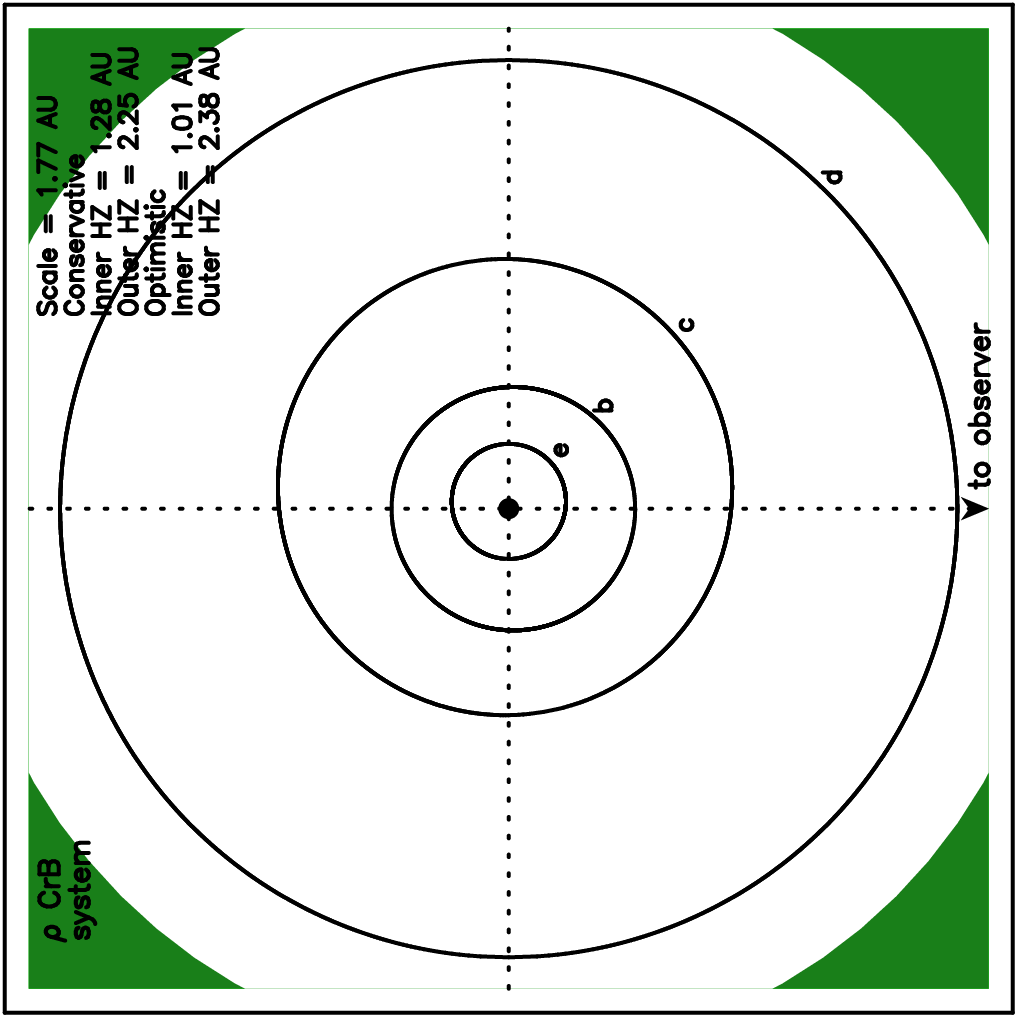}
  \caption{HZ and planetary orbits in the $\rho$~CrB system, where the
    orbits are labeled by planet designation. The inner edge of the
    OHZ is shown in green, and can be seen in the corners of the
    figure beyond the orbit of planet d. The scale of the figure is
    1.77~AU along each side.}
  \label{fig:hz}
\end{figure}

The $\rho$~CrB system consists of a central star and four known
orbiting planets. Given the brightness of the star ($V = 5.39$), there
are numerous stellar parameters available in the literature. We adopt
both the stellar and planetary parameters provided by
\citet{brewer2023}. The star has the following properties: mass
$M_\star = 0.95$~$M_\odot$, radius $R_\star = 1.34$~$R_\odot$,
effective temperature $T_\mathrm{eff} = 5817$~K, luminosity $L_\star =
1.82$~$L_\odot$, metallicity $\mathrm{[Fe/H]} = -0.20$~dex, and an age
of 10.2~Gyr. The orbits for the four known planets of the system are
relatively evenly spaced, and the planets have masses ranging from
super-Earth to Jovian. The properties of the planets are summarized in
Table~\ref{tab:planets}, and their orbits are depicted in a top-down
view of the system in Figure~\ref{fig:hz}, where the letter
designation of the planets are shown. The parameters shown in
Table~\ref{tab:planets} are the orbital period, $P$, semi-major axis,
$a$, eccentricity, $e$, argument of periastron, $\omega$, and the
minimum planetary mass, $M_p \sin i$.

\begin{deluxetable}{cccccc}
  \tablecolumns{6}
  \tablewidth{0pc}
  \tablecaption{\label{tab:planets} $\rho$~CrB planetary properties.}
  \tablehead{
    \colhead{Planet} &
    \colhead{$P$} &
    \colhead{$a$} &
    \colhead{$e$} &
    \colhead{$\omega$} &
    \colhead{$M_p \sin i$} \\
    \colhead{} &
    \colhead{(days)} &
    \colhead{(AU)} &
    \colhead{} &
    \colhead{($^\circ$)} &
    \colhead{($M_\oplus$)}
  }
  \startdata
  e &  12.949  & 0.1061 & 0.126 & 359.4  &   3.79 \\
  b &  39.8438 & 0.2245 & 0.038 & 269.64 & 347.4 \\
  c & 102.19   & 0.4206 & 0.096 &   9.7  &  28.2 \\
  d & 282.2    & 0.827  & 0.0   &   0.0  &  21.6
  \enddata
  \tablecomments{Planetary properties extracted from
    \citet{brewer2023}.}
\end{deluxetable}

Based on the stellar properties described above, we calculated the HZ
of the star using the methodology of
\citet{kopparapu2013a,kopparapu2014}. The traditional HZ, also
referred to as the conservative HZ (CHZ), is bounded by limits of
runaway and maximum greenhouse for an Earth analog. The HZ can be
empirically extended beyond the CHZ via the assumption that Venus and
Mars previously harbored surface liquid water, resulting in an
optimistic HZ (OHZ). The calculations of the CHZ and OHZ are described
in detail by \citet{kane2016c}. Since $\rho$~CrB is near the end of
its MS lifetime, the large radius and luminosity compared to solar
moves the HZ beyond the orbit of the outermost planet, and the inner
edge of the OHZ lies at $\sim$1.01~AU. The inner OHZ is shown as the
green regions in Figure~\ref{fig:hz}, the evolution of which will be
discussed further in Section~\ref{discussion}.

%%%%%%%%%%%%%%%%%%%%%%%%%%%%%%%%%%%%%%%%%%%%%%%%%%%%%%%%%%%%%%%%%%%%

\section{Stellar Evolution Model}
\label{mist}

To investigate the stellar evolution of $\rho$~CrB, we utilize the
MESA Isochrones \& Stellar Tracks (MIST; version 1.2) to calculate an
interpolated evolutionary track
\citep{paxton2011,paxton2013,paxton2015,choi2016,dotter2016,paxton2018,paxton2019}.
As described by \citet{choi2016}, MIST treats mass loss associated
with advanced stages of stellar evolution via the prescriptions
provided by \citet{reimers1975} and \citet{bloecker1995a}. Initial
bulk metallicities are computed by MIST assuming the protosolar
abundances provided by \citet{asplund2009}. Using the
\citet{brewer2023} stellar parameters of mass $M_\star =
0.95$~$M_\odot$ and metallicity $\mathrm{[Fe/H]} = -0.20$~dex (see
Section~\ref{arch}), the MIST algorithm estimated an initial helium
fraction of $Y_\mathrm{init} = 0.2627$. We set the initial stellar
surface velocity to 40\% of the critical (break-up) velocity,
$v/v_\mathrm{crit} = 0.4$. It is worth noting that, due to atomic
diffusion processes, the resulting evolutionary track is largely
unaffected by small ($< 0.1$~dex) changes in initial metallicity. It
is further of note that MIST tracks apply specifically to single-star
evolutionary models, and do not account for interactions with other
stars, such as the case of \citet{hon2023}. Since $\rho$~CrB is known
to be a single star, the MIST tracks are well-suited to the analysis
reported here.

\begin{figure*}
  \begin{center}
    \includegraphics[angle=270,width=16.0cm]{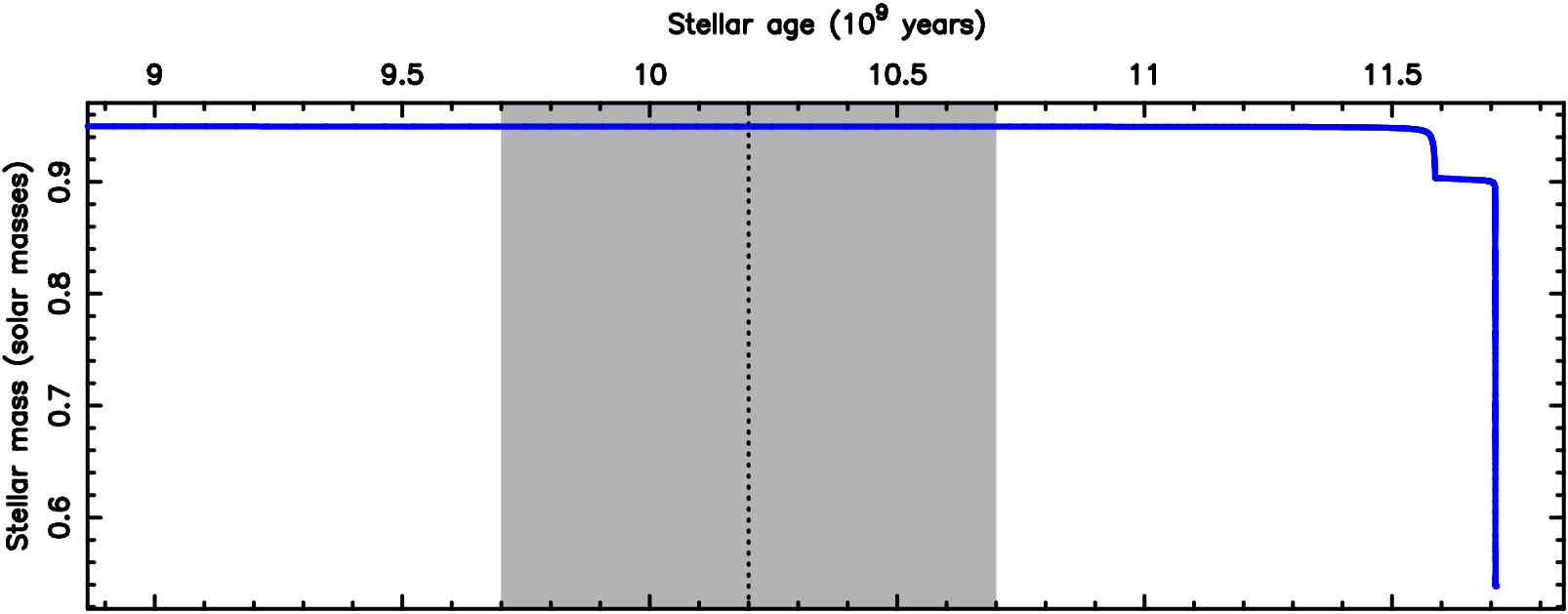} \\
    \includegraphics[angle=270,width=16.0cm]{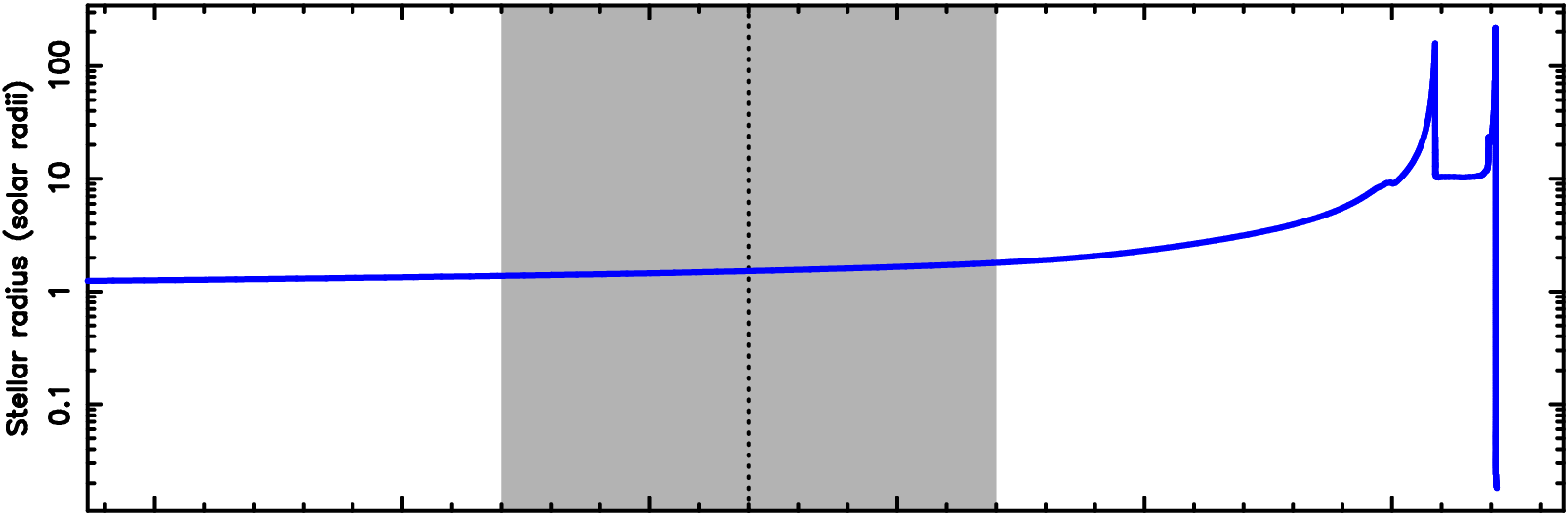} \\
    \includegraphics[angle=270,width=16.0cm]{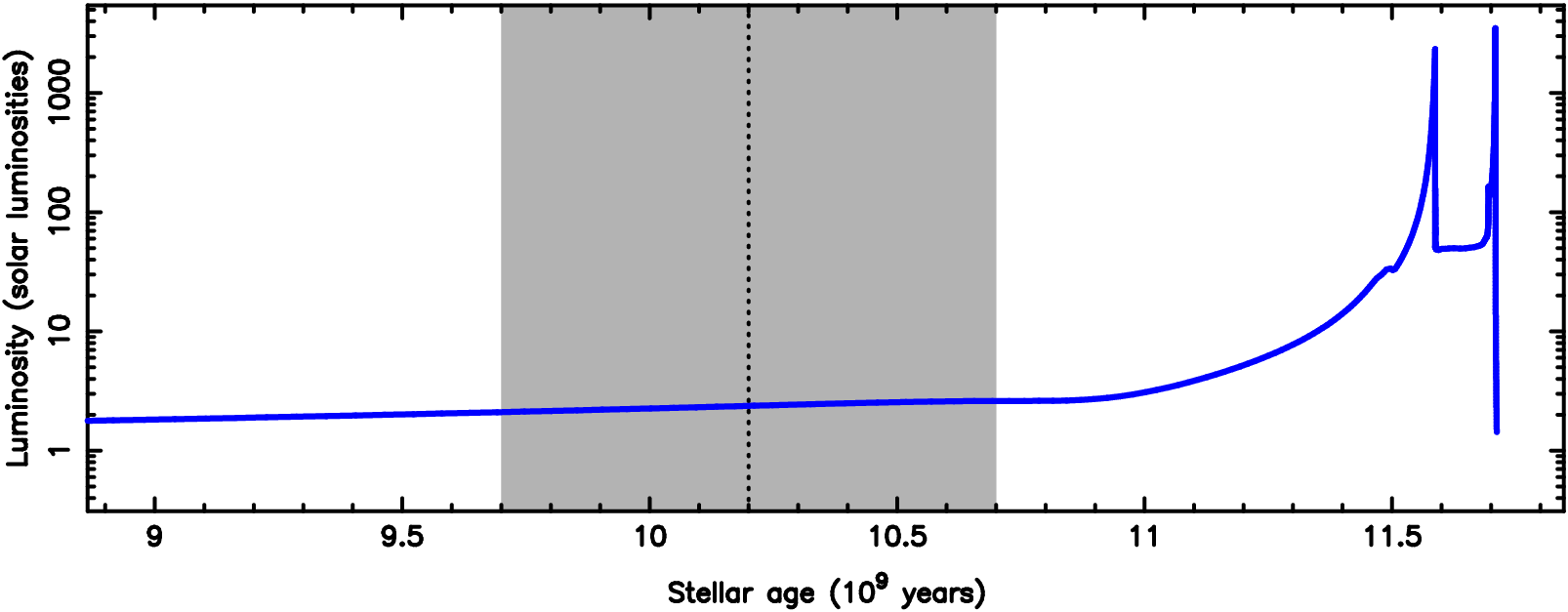}
  \end{center}
  \caption{The final predicted stages of MS evolution for $\rho$~CrB
    based on the MIST model described in Section~\ref{mist}, showing
    the evolution of stellar mass (top panel), radius (middle panel),
    and luminosity (bottom panel). The model depicts the transition
    through the RGB, horizontal branch, and AGB. The vertical dotted
    line indicates the current stellar age, and the gray shaded region
    represents the 1$\sigma$ uncertainty on that age.}
  \label{fig:mist}
\end{figure*}

The resulting stellar evolution model is shown in
Figure~\ref{fig:mist} for the final few billion years leading to the
progression of the star into the red-giant branch (RGB). Specifically,
Figure~\ref{fig:mist} shows the changes in stellar mass, radius, and
luminosity in solar units. The vertical dotted line indicates the age
estimate from \citep{brewer2023}, and the gray shaded region
represents the 1$\sigma$ uncertainty on the age. The sub-solar
metallicity of $\rho$~CrB results in an accelerated evolution of the
star into the RGB \citep{gehrig2023a}. The first peak in the stellar
radius/luminosity at $\sim$11.6~Gyr corresponds to the helium flash
and the transition into the horizontal branch
\citep{bloecker1995a,bloecker1995b}. The stellar radius decreases at
this transition, along with a $\sim$5\% mass loss. The horizontal
branch phase lasts for $\sim$$10^8$~years, beyond which the core
helium depletion again increases the radius as the star enters the
asymptotic giant branch (AGB). The MIST model predicts that the AGB
phase will peak at a stellar age of $\sim$11.7~Gyr, followed by a
shedding of the stellar envelope and dramatic mass loss. The
consequences of these stellar evolutionary processes for the known
planets will depend on their masses and semi-major axis relative to
the changing stellar radius.

%%%%%%%%%%%%%%%%%%%%%%%%%%%%%%%%%%%%%%%%%%%%%%%%%%%%%%%%%%%%%%%%%%%%

\section{Planetary Engulfment Prognosis}
\label{prog}

The expansion of a host star into the red giant phase can have a
variety of repercussions for the planets in the system. The engulfment
of planets can itself have several outcomes, depending on the specific
architecture of the system. Hydrodynamic simulations suggest that
engulfed planets may in-spiral over years or decades, and are
eventually destroyed either through evaporation or tidal disruption at
the Roche limit, puffing up the star in the process
\citep{staff2016b}. Some models indicate that sub-Jupiter mass planets
will not survive if initially located interior to 3--5~AU
\citep{villaver2007b}. Dynamical interactions between planets in the
system can rescue even smaller planets, as the stellar mass loss and
tidal effects can drive planetary orbits toward mean motion resonances
and increased separation from the host star \citep{ronco2020}. Such
surviving planets can remain long-term stable well beyond the RGB and
AGB phases of the star's evolution \citep{duncan1998a}. On the other
hand, planetary dynamics and tidal interactions may actually
precipitate planetary engulfment of close-in planets during the giant
star phase \citep{villaver2009}. These scenarios have been
investigated for several known exoplanet systems that are either
approaching \citep{rapoport2021}, presently in \citep{bear2011b}, or post 
\citep{charpinet2011b} the RGB/AGB phases of the host star.

\begin{figure*}
  \begin{center}
    \includegraphics[angle=270,width=16.0cm]{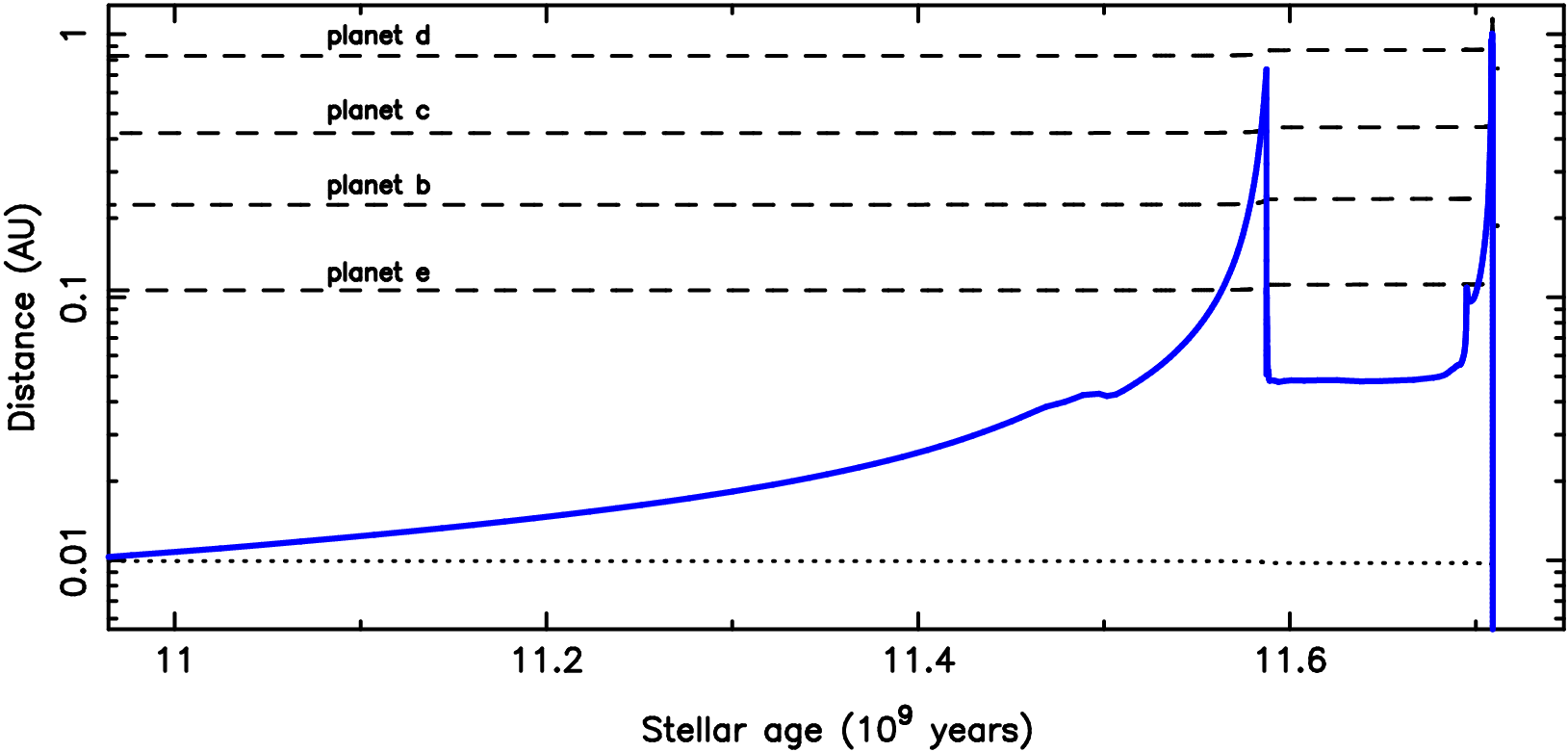} \\
    \includegraphics[angle=270,width=16.0cm]{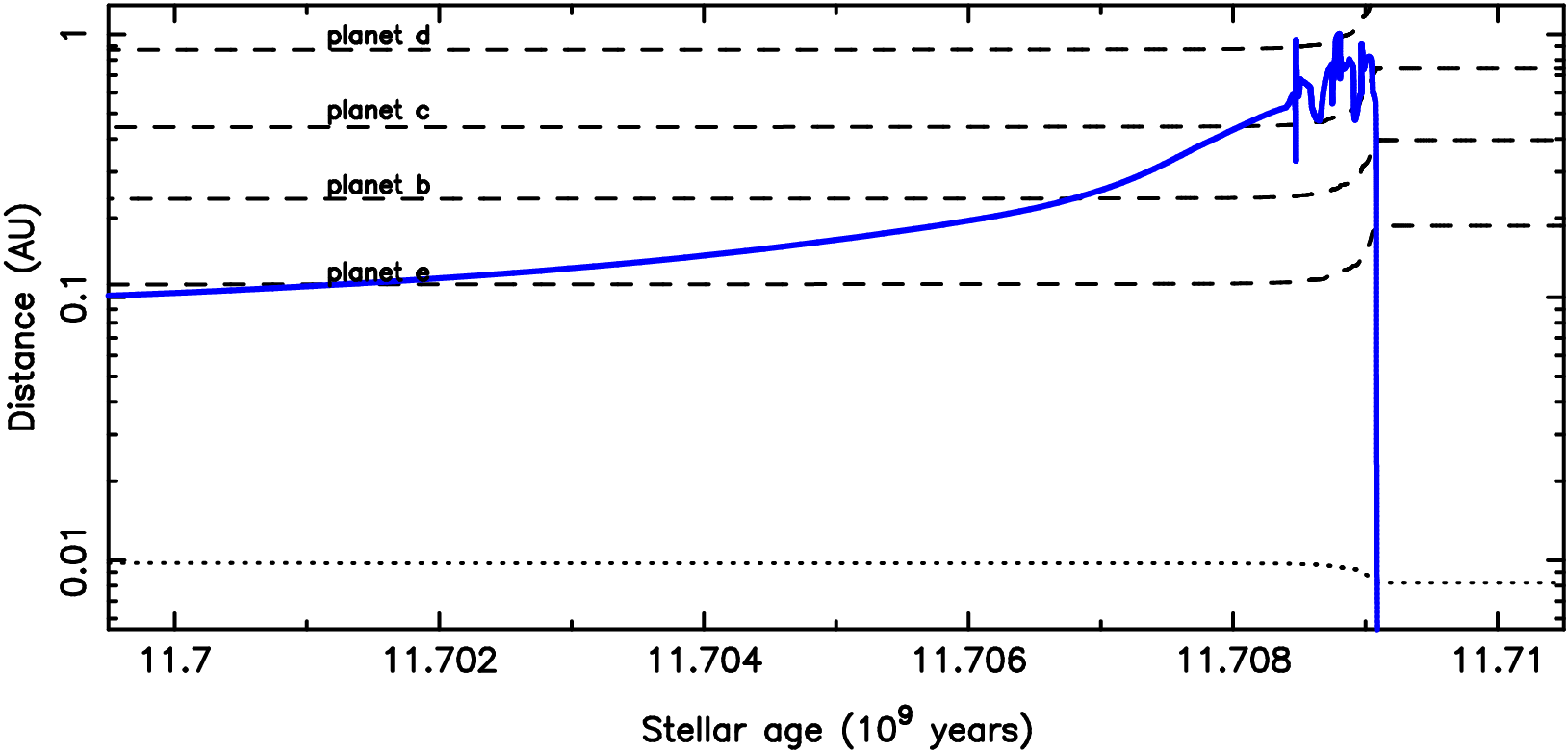}
  \end{center}
  \caption{Distance from the host star in AU for the stellar evolution
    off the MS over $7 \times 10^8$~years (top panel) and $10^7$~years
    (bottom panel). The radius of the star is shown as a blue line,
    and the dashed lines show the semi-major axes of the known
    planets. The location of the fluid body Roche limit of the star is
    indicated by the dotted line.}
  \label{fig:prog}
\end{figure*}

To determine the effect of the stellar evolution model described in
Section~\ref{mist} on the $\rho$~CrB system planets, we overlaid the
evolving stellar properties against the orbits of the planets. A $7
\times 10^8$~year segment of the stellar radius evolution is shown in
the top panel of Figure~\ref{fig:prog} as a blue line and in units of
AU. The semi-major axes of the planets are indicated by the horizontal
dashed lines, and have been recalculated at each epoch to incorporate
stellar mass loss and conservation of angular momentum for each
orbit. The dotted line represents the location of the fluid body Roche
limit for the star which, despite the dramatic change in stellar
radius, remains relatively constant since it is dominated by the
stellar mass. The bottom panel of Figure~\ref{fig:prog} further zooms
in on the final $10^7$~years at the end of the AGB phase before the
transition to the white dwarf phase. The stellar radius fluctuations
visible in the range 11.708--11.709~Gyrs are the signatures of thermal
pulses caused by shell hydrogen and helium fusion within the giant
star.

As the star swells in size during the RGB phase, the inner planets of
e, b, and c are engulfed by the star at stellar ages of 11.5630,
11.5785, and 11.5846~Gyrs, respectively. The star achieves a maximum
RGB size of $\sim$158~$R_\odot$ (0.736~AU) at an age of 11.5874~Gyrs,
at which point the star encompasses all planets except for the outer
planet, d. After the helium flash, the star decreases in size as it
enters the horizontal branch, and plateaus at a minimum size of
$\sim$10.4~$R_\odot$ (0.048~AU). As the helium fuel is depleted, the
stellar radius increases again into the AGB, exceeding its RGB maximum
size, and reaching $\sim$216~$R_\odot$ (1.005~AU) at
11.7088~Gyrs. This period during the AGB is also characterized by the
aforementioned radius fluctuations due to shell fusion burning along
with significant mass loss. Conservation of angular momentum causes
the semi-major axis of planet d to increase to 0.965~AU, but this is
insufficient to save the planet from engulfment. Planet d is swallowed
by the star during the AGB phase at 11.7085~Gyrs. It remains 0.05~AU
interior to the star for about a thousand years, after which the star
contracts again. At 11.7088~Gyrs, the star engulfs planet d again,
where it is once again 0.05~AU interior to the star for several
thousand years. If planet d can survive this phase, it will settle
into a 1.459~AU orbit as the star rapidly transitions into a white
dwarf. Thus, all four of the known planets will likely be engulfed by
their star within the next 1.0--1.5 billion years.

%%%%%%%%%%%%%%%%%%%%%%%%%%%%%%%%%%%%%%%%%%%%%%%%%%%%%%%%%%%%%%%%%%%%

\section{Discussion}
\label{discussion}

Although all of the planets will enter the stellar atmosphere of
$\rho$~CrB, their individual prognoses vary considerably. Planet e is
likely terrestrial in nature and, given it is the first to be engulfed
deep into the star, an evaporation scenario may be the most probable
outcome. However, planet b is more massive than Jupiter, and the
immersion into the stellar envelope will result in viscous drag and
in-spiral, ending with tidal disruption at the Roche limit
\citep{staff2016a,oconnor2023a}, shown to remain relatively stable at
$\sim$0.01~AU in Figure~\ref{fig:prog}. The accretion of planetary
material at the base of the convective envelope can cause a further
increase in stellar size \citep{siess1999b}, which is not taken into
account in our model. If indeed such an additional stellar radius
increase occurs, then the engulfment of planet c may take place
earlier than that stated in Section~\ref{prog}, and may also result in
the engulfment of planet d whilst the star is still on the RGB. Both
planet c and d are similar in mass to Neptune and will therefore
suffer substantial evaporation over an orbital in-spiral scenario. Our
model further did not include the effects of orbital dynamics, which
has the potential to cause planet d to migrate further outward and
possibly escape engulfment \citep{ronco2020}. Such planetary
interactions are particularly important in the case of resonance
crossing events \citep{kane2023a}, such as is predicted for the solar
system post-MS evolution \citep{zink2020c}. Since the inner planets of
$\rho$~CrB are engulfed prior to the AGB phase, it is unlikely that
orbital dynamics will play a major role in the system during and after
the stellar mass loss.

For cases where planets are consumed by the host star, several
mechanisms have been invoked that may yield observable signatures of
such consumption events
\citep{siess1999a,siess1999b,stephan2020a,behmard2023a}. These
signatures can include enhanced lithium abundance
\citep{aguileragomez2016a,sevilla2022}, stronger magnetic fields
\citep{privitera2016b}, and faster rotation rates
\citep{privitera2016c}. The latter of these signatures may also be
connected to enhanced mass loss on the RGB
\citep{soker1998c,bear2011a}, further improving the survival rates for
outer planets in the system through angular momentum
conservation. Thus far, observational evidence for planetary
engulfment signatures has remained relatively sparse, suggesting that
either engulfment scenarios are rarer than expected, or that signature
detection is more challenging than anticipated \citep{behmard2023c}.

Another signature of planet engulfment may be found through a careful
examination of white dwarfs \citep{nelemans1998,mustill2012b}. White
dwarf planetary systems are often assumed to be quite prevalent
\citep{zuckerman2010} with orbital architectures that are
intrinsically linked to the prior evolution of the progenitor
\citep{debes2002a}. Stellar evolution and planetary engulfment is a
possible interpretation for white dwarf pollution
\citep{frewen2014,petrovich2017a}, which can also be caused by more
recent planetary accretion events \citep{gansicke2019}. Numerous giant
planets have been discovered orbiting white dwarfs
\citep{vanderburg2020b,blackman2021b}, including those whose presence
have been interpreted within the context of RGB survival
\citep{lagos2021,merlov2021}.

White dwarf planetary systems are also of interest with respect to the
prospects for habitable planets that may be present. Given the
relatively large transit depth, searches have been proposed using the
transit method that specifically target terrestrial planets in the HZ
\citep{agol2011}. White dwarf HZ planets face additional challenges
toward maintaining long-term temperate surface conditions, such as the
cooling of the star \citep{barnes2013b} and tidal effects
\citep{becker2023}. These HZ planets were almost certainly not in the
HZ during the MS or RGB/AGB phases of the progenitor, particularly
given the mass loss that occurs during the AGB. Section~\ref{arch} and
Figure~\ref{fig:hz} provide the current HZ boundaries for $\rho$~CrB,
where the OHZ extends in the range 1.01--2.38~AU. No planets have yet
been detected in the HZ, and the predicted RV semi-amplitude for an
Earth-mass planet is 9.1~cm/s and 5.9~cm/s at the inner and outer
edges of the OHZ, respectively (assuming that the orbit is close to
edge-on). As shown in Section~\ref{prog}, the star will achieve a
maximum size of 1.005~AU, almost reaching the inner edge of the
OHZ. Due to stellar mass loss during the AGB, a planet that presently
lies in the middle of the OHZ (1.70~AU) will move outward to a
semi-major axis of $\sim$3.0~AU. During the horizontal branch, the
bottom panel of Figure~\ref{fig:mist} shows that the luminosity will
be two orders of magnitude larger than it is currently. The HZ
boundaries defined by \citet{kopparapu2013a,kopparapu2014} scale with
$L_\star^{0.5}$, and so they will be $\sim$10 times larger during the
horizontal branch, which is beyond the estimated new semi-major axis
of $\sim$3.0~AU. On the other hand, the new semi-major axis will place
the planet well exterior to the white dwarf HZ. Thus, neither the
possibly surviving planet d or a hypothetical planet currently
residing in the HZ will be present in the HZ during either the RGB/AGB
or white dwarf phases.

%%%%%%%%%%%%%%%%%%%%%%%%%%%%%%%%%%%%%%%%%%%%%%%%%%%%%%%%%%%%%%%%%%%%

\section{Conclusions}
\label{conclusions}

The evolution of stars through their progression on the MS, expansion
into a giant star, and then final contraction into a white dwarf, has
profound consequences for the orbiting planets. The case studied here,
that of the $\rho$~CrB system, is particularly interesting due to the
brightness and late age of the star, and the diverse planetary system
that is currently known to extend to distances of $\sim$0.83~AU from
the host. The sub-solar metallicity of the star truncates the MS
lifetime, and our model predicts that it will reach the end of the AGB
within 1.0--1.5~billion years. Given the masses and semi-major axes of
the four known planets, we predict that planet e will evaporate within
the stellar atmosphere, planet b will in-spiral and be tidally
disrupted, potentially further inflating the star, and planet c will
be evaporated within the stellar atmosphere. The fate of the outermost
planet, planet d, remains uncertain but will likely be evaporated
within the star during the end of the AGB. If it is able to escape
engulfment during stellar mass loss, it may remain in orbit around the
white dwarf at a separation of $\sim$1.5~AU. For HZ planets that may
be present in the system but below current detection sensitivity, they
will survive the stellar evolution but be interior to the HZ inner
edge during the RGB/AGB phase and exterior to the HZ outer edge during
the white dwarf phase.

Continued exoplanet monitoring of nearby bright stars that are
approaching the end of their MS lifetime will provide additional
opportunities to study the effects of stellar evolution on planetary
systems. Orbital dynamics may play an important role in the
survivability of inner system planets, and thus the detection of outer
giant planets can provide essential keys in understand the full
prognosis for a given system. In that regard, the addition of
astromtric measurements from Gaia \citep{brown2021} for the existing
RV data of nearby systems will greatly improve the detection
sensitivity in the outer regions of exoplanetary systems
\citep{wright2009b}. The further merging in of direct imaging data
will provide additional constraints on the system architecture
\citep{brandt2019,kane2019b}, as well reflected and emission spectra
for any detected planets \citep{stark2020,li2021a,saxena2021}.
Connecting such compositional information with stellar abundances may
yield critical insight regarding signatures of post-MS planet
engulfment.

%%%%%%%%%%%%%%%%%%%%%%%%%%%%%%%%%%%%%%%%%%%%%%%%%%%%%%%%%%%%%%%%%%%%

\section*{Acknowledgements}

The author is grateful to the referee for their valuable feedback on
the manuscript. This research has made use of the Habitable Zone
Gallery at hzgallery.org. The results reported herein benefited from
collaborations and/or information exchange within NASA's Nexus for
Exoplanet System Science (NExSS) research coordination network
sponsored by NASA's Science Mission Directorate.

%%%%%%%%%%%%%%%%%%%%%%%%%%%%%%%%%%%%%%%%%%%%%%%%%%%%%%%%%%%%%%%%%%%%

%\bibliographystyle{aasjournal}
%\bibliography{/data/skane/latex/styles/references}

%%%%%%%%%%%%%%%%%%%%%%%%%%%%%%%%%%%%%%%%%%%%%%%%%%%%%%%%%%%%%%%%%%%%

\end{document}